\documentclass{80SA}
\usepackage{times}

\title{Variational Monte Carlo Study of the Kondo Necklace Model with Geometrical Frustration}

\author{Yukitoshi Motome, Kyoya Nakamikawa, Youhei Yamaji, and Masafumi Udagawa
}
\inst{Department of Applied Physics, University of Tokyo, Hongo, Tokyo 113-8656, Japan
}
\abst{We investigate the ground state of the Kondo necklace model 
on geometrically-frustrated lattices by the variational Monte Carlo simulation. 
To explore the possibility of a partially-ordered phase, 
we employ an extension of the Yosida-type wave function as a variational state, 
which can describe a coexistence of spin-singlet formation due to the Kondo coupling 
and magnetic ordering by the Ruderman-Kittel-Kasuya-Yosida interaction. 
We show the benchmark of the numerical simulation to demonstrate the high precision 
brought by the optimization of a large number of variational parameters. 
We discuss the ground-state phase diagram for the model 
on the kagome lattice in comparison with that for the triangular-lattice case. 
}

\kword{geometrical frustration, Kondo coupling, RKKY interaction, Kondo necklace model, partial order, variational Monte Carlo method, triangular lattice, kagome lattice}

\begin{document}
\maketitle

\section{Introduction}

Competition between different electronic phases is a key to understand 
the variety of fascinating phenomena in strongly-correlated electron systems. 
An interesting situation shows up when different phases compete with each other 
at zero temperature ($T$) to result in a quantum critical point (QCP). 
QCP is a source of many exotic properties, 
such as a non-Fermi-liquid behavior and some emergent phases. 

A typical example of QCP and related phenomena is found in the Kondo lattice systems 
such as $f$-electron systems~\cite{Hewson}. 
In the Kondo lattice systems, competition originates in two different types of interactions; 
one is the Kondo coupling between conduction electrons and localized moments, 
and the other is the Ruderman-Kittel-Kasuya-Yosida (RKKY) interaction which is 
an effective magnetic coupling between localized moments mediated by conduction electrons. 
The former induces the formation of the Kondo spin-singlet~\cite{Kondo1964}, 
whereas the latter tends to stabilize a magnetic long-range ordering~\cite{Ruderman1954}. 
The competition leads to QCP between a Fermi liquid state and 
a magnetically-ordered (MO) state, which has been argued as a key concept 
for understanding rich behaviors in the Kondo lattice systems~\cite{Doniach1977}. 

Geometrical frustration is another important concept in strongly-correlated electron systems. 
Frustration suppresses conventional magnetic orderings and
opens up the possibility to have exotic phases, 
such as fluctuation-induced phases by the order-from-disorder mechanism~\cite{Villain1980}. 
One of such intriguing phenomena is a partial order:~\cite{Mekata1977} 
The system is separated into magnetic and nonmagnetic sublattices 
to relieve the frustration; 
namely, a magnetic ordering appears on a sublattice of the system 
with leaving the remaining sublattice nonmagnetic. 
The partial ordered states are in most cases stabilized by thermal fluctuations, 
i.e., by a gain of the entropy. 

Recently, a different type of partial order was explored in the Kondo lattice system 
by the authors~\cite{Motome2010}. 
The partial order is not entropic-driven but a new quantum state 
stabilized at zero $T$ under geometrical frustration 
in the competing region between the Kondo coupling and the RKKY interaction. 
The partial ordered phase consists of a magnetically-ordered unfrustrated network in a sublattice 
and nonmagnetic spin-singlet sites with Kondo screening in the remaining sublattice. 
We call this state the partial Kondo screening (PKS) state. 
The phase diagram was studied by the variational Monte Carlo (VMC) method, and 
it was shown that PKS is further stabilized by quantum fluctuations and the spin anisotropy. 

In this contribution, we report the numerical details of the VMC calculations. 
By showing the extended benchmark results with detailed conditions of the calculations, 
we demonstrate that VMC gives the ground-state energy in high precision, 
typically within 1-3\% relative errors. 
We also show a possible ground-state phase diagram for 
the Kondo necklace model (KNM) on the two-dimensional kagome lattice. 
We discuss the stability of PKS in comparison with the results for the triangular-lattice case.

\section{Model and Method}

\begin{figure}
\begin{center}
 \includegraphics[width=.47\textwidth,clip]{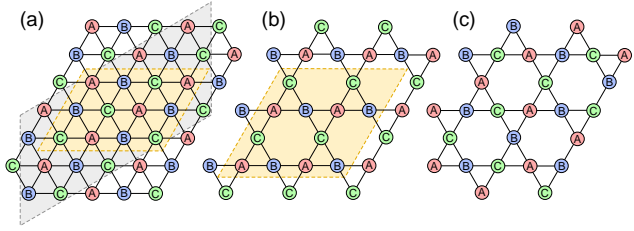}
\end{center}
\caption{(Color online) 
Schematic pictures of three-sublattice orderings on 
(a) the triangular lattice and (b), (c) the kagome lattice. 
(b) is the $\vec{q}=0$ ordering and (c) is the $\sqrt{3}\times\sqrt{3}$ ordering. 
12- and 24-site clusters used in the present calculations are also shown~\cite{note}.}
\label{fig:lattices}
\end{figure}

In the present study, we consider KNM 
which has been studied as a mimic of the half-filling case of 
the Kondo lattice model~\cite{Doniach1977}.
The Hamiltonian is given by
\begin{equation}
{\cal H} = W \sum_{\langle ij \rangle} \vec{\tau}_i \cdot \vec{\tau}_j 
+ J \sum_i \vec{\tau}_i \cdot \vec{S}_i 
+ I_z \sum_{\langle ij \rangle} S_i^z S_j^z, 
\label{eq:H_KNM}
\end{equation}
where the first term describes the AF interaction 
between conduction electron spins $\vec{\tau}_i$ 
and the second term represents the Kondo coupling between $\vec{\tau}_i$ and 
the localized spin $\vec{S}_i$. 
The former tends to stabilize a magnetic ordering, and 
the latter favors the Kondo spin-liquid (KSL) state 
in which all the sites participate in the local Kondo singlet formation. 
Hence, in the unfrustrated cases, this model describes QCP between MO and KSL, 
instead of MO and a Fermi liquid state, because of the half filling~\cite{Doniach1977}. 
For simplicity, we consider $S=1/2$ spins for both $\vec{\tau}_i$ and $\vec{S}_i$. 
In Eq.~(\ref{eq:H_KNM}), we extend the model by adding the last term, 
the AF Ising interaction between the localized spins 
($S_i^z$ is the $z$ component of $\vec{S}_i$), 
in order to incorporate the magnetic anisotropy often seen in real materials. 
We consider the model on two different frustrated lattice structures in two dimensions, 
the triangular lattice and the kagome lattice, shown in Fig.~\ref{fig:lattices}. 
The sums $\langle ij \rangle$ in Eq.~(\ref{eq:H_KNM}) are taken over 
the nearest-neighbor pairs on these lattices. 
Hereafter we take $W=1$ as the energy unit.

We study the ground state of the model in Eq.~(\ref{eq:H_KNM}) 
by the VMC method. 
To explore the PKS state in which nonmagnetic spin singlet and 
magnetic ordering coexist, we here consider an extension of 
the Yosida-type wave function~\cite{Shiba1990} 
as the variational wave function. 
The functional form is given by
\begin{equation}
|\psi \rangle = P_{\rm{G}} {\cal L}^{S=0} {\cal L}^{K=0} |\phi_{\rm{pair}} \rangle. 
\label{eq:psi}
\end{equation}
Here $|\phi_{\rm{pair}} \rangle$ is a generalized BCS wave function given by 
\begin{equation}
|\phi_{\rm{pair}} \rangle = 
\Big( \sum_{\ell,m=1}^{2N} f_{\ell m} a_{\ell \uparrow}^\dagger a_{m \downarrow}^\dagger \Big)^N |0 \rangle,
\label{eq:BCS}
\end{equation}
where $f_{\ell m}$ are the variational parameters, 
$N$ is the number of lattice sites, and $|0 \rangle$ is a vacuum. 
The pair creations with fermion operators $a_{\ell\sigma}^\dagger$ include 
both conduction and localized spin degrees of freedom 
(the sums run over the lattice sites with distinguishing them); 
the electron numbers are fixed to be one in each degrees of freedom 
by the Gutzwiller projection operator $P_{\rm{G}}$ in Eq.~(\ref{eq:psi}). 
${\cal L}^{S=0}$ and ${\cal L}^{K=0}$ are the quantum-number projection operators 
for the total spin singlet and the total momentum zero, respectively. 
We optimize a large number of variational parameters
by using the stochastic reconfiguration~\cite{Sorella1998} and 
enforce the quantum-number projections, 
by following the method in Ref.~10. 

\section{Benchmark}

In this section, we show the benchmark results of our numerical calculations. 
In the following, we explore the ground states with three-sublattice ordering 
in two different ways. 
One is to impose the momentum projection ${\cal L}^{K=0}$ only for the same sublattices 
with taking account of all possible $f_{lm}$. 
The other is to impose the three-sublattice translational symmetry on $f_{lm}$, 
which largely reduces the number of $f_{lm}$ required in the calculations. 
We take the 12- and 24-site clusters shown in Fig.~\ref{fig:lattices} 
under the periodic boundary conditions compatible with the three-sublattice orderings. 
We apply the spin-singlet projection ${\cal L}^{S=0}$ only for the case with $I_z=0$ hereafter. 
We typically perform $300$-$1000$ stochastic reconfiguration steps 
with $1600$-$6000$ MC samplings in the following calculations. 
The benchmark is shown for the parameter range where the three-sublattice orders 
(the MO and PKS states) are obtained. 

First we present the results for the triangular-lattice case. 
Table~\ref{table:triangle12} shows the total ground-state energy for the 12-site system 
obtained by VMC in comparison with the result by the exact diagonalization. 
The VMC results show the ground state energy 
by the first method, i.e., by the full optimization of $f_{lm}$ with ${\cal L}^{K=0}$. 
As shown in the table, VMC provides the energy very close to the exact value 
in a wide range of parameters $J$ and $I_z$. 
The relative errors are typically $\sim 1$-$3$\%. 

Table~\ref{table:kagome12} presents similar comparisons for the 12-site kagome-lattice models. 
Note that the system with $N=12$ under the periodic boundary conditions 
accommodates only the $\vec{q}=0$ type three-sublattice ordering, 
as shown in Fig.~\ref{fig:lattices}(b). 
As presented in the table, VMC gives the ground-state energy 
in high accuracy also in this kagome-lattice case, typically within $1$-$2$\% relative errors. 

Finally, we compare the two methods, namely, 
the full optimization of $f_{lm}$ with ${\cal L}^{K=0}$ and 
the optimization of a reduced set of $f_{lm}$ with three-sublattice periodicity. 
The results for the 24-site triangular-lattice models are shown in Table~\ref{table:triangle24}. 
For the 24 site, the number of the variational parameters $f_{ij}$ is 
$2304$ for the full optimization, 
whereas it is reduced to $288$ when the three-sublattice periodicity is imposed. 
The results show that the reduced set still gives good precision,~\cite{note2} 
in particular, for larger $J$ where spatial correlations become local. 
This encourages us to extend the calculations for larger cluster sizes 
by using the reduced set of $f_{lm}$ which greatly reduces the CPU cost 
compared to the full optimization. 
The extension will be reported elsewhere. 

\begin{table}[t]
\caption{Ground-state energy for KNM on the triangular lattice with $N=12$. 
Comparison between the VMC results 
and the exact diagonalization results are shown with relative errors.
The numbers in the parentheses indicates the statistical errors in the last digit. 
The upper table is for the results at $I_z=0$ and the lower at $I_z=0.4$. 
\label{table:triangle12}
}
\begin{tabular}{c|cccc}
$J$ & $0.2$ & $0.4$ & $0.6$ & $0.8$ \\ 
\hline
VMC & $-7.38(1)$ & $-7.98(1)$ & $-8.93(1)$ & $-10.16(1)$ \\
exact & $-7.446$ & $-8.061$ & $ -9.025$ & $-10.232$ \\
error & $0.009(1)$ & $0.010(1)$ & $0.011(1)$ & $0.007(1)$ 
\end{tabular}
\begin{tabular}{c|cccc}
$J$ & $0.2$ & $0.4$ & $0.6$ & $0.8$ 
\\
\hline
VMC & $-8.49(1)$ & $-8.92(1)$ & $-9.72(1)$ & $-10.77(1)$ \\
exact & $-8.602$ & $-9.103$ & $ -9.957$ & $-11.059$ \\
error & $0.013(1)$ & $0.020(1)$ & $0.024(1)$ & $0.026(1)$ 
\end{tabular}
\end{table}

\begin{table}
\vspace*{-6mm}
\caption{
Ground-state energy for KNM on the kagome lattice with $N=12$. 
The upper table is for $I_z=0$, while the lower for $I_z=0.4$. 
\label{table:kagome12}
}
\begin{tabular}{c|cccc}
$J$ & $0.14$ & $0.16$ & $0.18$ & $0.20$ \\
\hline
VMC & $-5.55(1)$ & $-5.58(1)$ & $-5.67(1)$ & $-5.76(1)$ \\
exact & $-5.578$ & $-5.642$ & $ -5.717$ & $-5.800$ \\
error & $0.005(2)$ & $0.011(2)$ & $0.008(2)$ & $0.007(2)$ 
\end{tabular}
\begin{tabular}{c|cccc}
$J$ & $0.14$ & $0.16$ & $0.18$ & $0.20$ \\
\hline
VMC & $-6.30(1)$ & $-6.33(1)$ & $-6.36(1)$ & $-6.40(1)$ \\
exact & $-6.349$ & $-6.403$ & $ -6.469$ & $-6.544$ \\
error & $0.008(2)$ & $0.011(2)$ & $0.017(2)$ & $0.022(2)$ 
\end{tabular}
\end{table}

\begin{table}[t]
\caption{
VMC ground-state energy for KNM on the triangular lattice with $N=24$ sites. 
The results show the comparison between the optimization of all $f_{lm}$ and 
of only three-sublattice symmetry-allowed $f_{lm}$. 
Smaller errors in the former cases are owing to the quantum projection ${\cal L}^{K=0}$. 
The upper table shows the results at $I_z=0$, while the lower at $I_z=0.4$. 
\label{table:triangle24}
}
\begin{tabular}{c|cccc}
$J$ & $0.2$ & $0.4$ & $0.6$ & $0.8$ \\
\hline
all $f_{lm}$ & -14.20(7) & -15.15(4) & -17.03(4) & -20.04(8) \\
3-sub $f_{lm}$ & -12.2(3) & -15.0(2) & -17.1(3) & -19.1(3) 
\end{tabular}
\begin{tabular}{c|cccc}
$J$ & $0.2$ & $0.4$ & $0.6$ & $0.8$ \\
\hline
all $f_{lm}$ & -16.62(3) & -17.60(4) & -19.31(3) & -21.49(7) \\
3-sub $f_{lm}$ & -14.6(3) & -16.8(1) & -18.6(5) & -20.9(2) 
\end{tabular}
\end{table}

\section{Results for the Kagome Lattice Case}

\begin{figure}[t]
\begin{center}
 \includegraphics[width=.42\textwidth,clip]{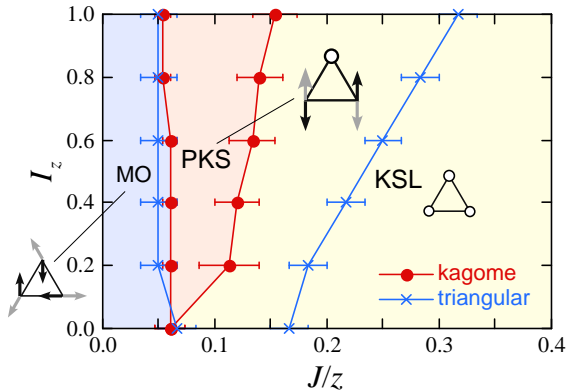}
\end{center}
\caption{(Color online) 
Phase diagram deduced from the VMC results for KNM on the kagome lattice with $N=12$ sites. 
The data for the triangular-lattice case with $N=12$ are taken from Ref.~7. 
Schematic pictures for the spin state in each phase are shown, in which the grey (black) arrows represent $\vec{\tau}_i$ ($\vec{S}_i$) and the circles denote the Kondo singlets. 
}
\label{fig:kagome_phase}
\end{figure}

Here we present the VMC results for KNM on the kagome lattice, 
in comparison with those for the triangular-lattice case 
previously reported in Ref.~7. 
In the kagome case, even when considering only the three-sublattice ordering, 
there are different ordering patterns, such as $\vec{q}=0$ or $\sqrt{3}\times\sqrt{3}$ 
[Figs.~\ref{fig:lattices}(b) and (c)]. 
This is because of the corner-sharing network of triangles. 
Since it is difficult to analyze the ground state with allowing all such possibilities 
in VMC 
with increasing the system size $N$, 
here we limit ourselves to a small size cluster $N=12$ under the periodic boundary conditions, 
which accommodates only the solutions compatible with the $\vec{q}=0$ type order. 
The following calculations are done by the full optimization of all $f_{lm}$ 
with imposing ${\cal L}^{K=0}$ for the same sublattices. 
We map out the ground-state phase diagram from the behaviors of 
the onsite spin correlations $\langle \vec{\tau}_i \cdot \vec{S}_i \rangle$ 
as well as the intersite ones $\langle \vec{S}_i \cdot \vec{S}_j \rangle$, 
following the manner in our previous study for the triangular-lattice models~\cite{Motome2010}. 

Figure~\ref{fig:kagome_phase} shows the results. 
For comparison, the results for KNM on the triangular lattice with $N=12$ are also presented. 
Note that the horizontal axis denotes $J$ renormalized by the coordination number $z$ 
(the number of nearest neighbors) to facilitate 	the comparison. 
We obtain a similar sequence of three different phases as in the triangular-lattice case, 
i.e., MO, PKS, and KSL phases, with increasing $J$. 
However, we observe that the PKS phase is largely suppressed in the kagome-lattice case, 
and that the KSL phase is relatively stabilized in the large $J$ region. 
In particular, we do not see a clear evidence of PKS at $I_z=0$, 
whereas the triangular-lattice case exhibits a finite window of $J$ for PKS down to $I_z=0$. 
The results suggest that PKS is relatively unstable 
in the kagome-lattice model compared to the triangular-lattice model. 
The weak tendency to PKS is reasonable 
since the magnetic sublattice in PKS has a relatively weak network 
in the kagome-lattice case compared to the triangular-lattice case: 
it forms a two-dimensional honeycomb network in the triangular-lattice case~\cite{Motome2010}, 
whereas it consists of one-dimensional chains 
in the $\vec{q}=0$ ordering in the kagome-lattice case. 
The tendency is also consistent with the observation in the limit of $I_z \gg J$ ($\gg W$) 
where the system can be mapped onto 
an effective transverse-field Ising model~\cite{Motome2010}: 
For the transverse-field Ising model, a three-sublattice partial order is suggested 
to be stabilized in the triangular-lattice case, 
whereas it is not in the kagome case~\cite{Moessner2001}. 

Even though we believe that the results capture the correct tendency, 
it is unclear whether the PKS state survives in the kagome-lattice model 
in the thermodynamic limit $N \to \infty$. 
Especially, it will be important to take account of 
all other possibilities of the three-sublattice ordering patterns. 
The most exotic possibility is a spin-liquid state composed of 
the PKS three-site units (two sites bear magnetic moments, 
while the remaining one participates in the Kondo-singlet formation): 
In the kagome lattice, the triangle units are connected with sharing only their corners, 
and hence, it is possible to have a magnetically-disordered state 
by connecting the PKS three-site units. 
In the extreme case, such state bears macroscopic degeneracy. 
It is interesting to ask whether the degeneracy is lifted 
through quantum order-from-disorder. 
These interesting issues are left for future study.

\section{Summary}

We have investigated the ground state of frustrated Kondo necklace model 
by the variational Monte Carlo calculations. 
We have shown the detailed benchmark for both the triangular and kagome lattice cases, 
which demonstrates 
that our simulation employing a generalized Yosida-type wave function 
gives highly-precise ground-state energy in a wide parameter range. 
We have also deduced the phase diagram for the kagome-lattice model. 
The comparison with the triangular-lattice case revealed 
the reduced tendency toward the partial Kondo screening in the kagome-lattice case.

\section*{Acknowledgment}
The authors thank D. Tahara for the use of his VMC code and useful comments, 
and T. Misawa for enlightening discussions. 
This work was supported by KAKENHI (Nos. 19052008 and 21340090), 
Global COE Program ``the Physical Sciences Frontier", 
and by the Next Generation Super Computing Project, Nanoscience Program, MEXT, Japan.
A part of the calculations were done by using TITPACK Ver.2 by H. Nishimori.


\begin{thebibliography}{9}
\bibitem{Hewson} A. C. Hewson, {\it `The Kondo Problem to Heavy Fermions'} (Cambridge Univ. Press, Cambridge, 1993), and references therein.
\bibitem{Kondo1964} J.\ Kondo: Prog. Theor. Phys. {\bf 32} (1964) 37; K. Yosida: Phys. Rev. {\bf 147} (1966) 223; P. W. Anderson: J. Phys. {\bf C3} (1970) 2436; Ph. Nozi${\rm \grave{e}}$res: J. Low. Phys. {\bf 17} (1974) 31.
\bibitem{Ruderman1954} M. A. Ruderman and C. Kittel: Phys. Rev. {\bf 96} (1954) 99; T. Kasuya: Prog. Theor. Phys. {\bf 16} (1956) 45; K. Yosida: Phys. Rev. {\bf 106} (1957) 893. 
\bibitem{Doniach1977} S. Doniach: Physica {\bf 91B} (1977) 231.
\bibitem{Villain1980} J. Villain, R. Bidaux, J. P. Carton, and R. Conte: J. Phys. {\bf 41} (1980) 1263. 
\bibitem{Mekata1977} M. Mekata: J. Phys. Soc. Jpn. {\bf 42} (1977) 76. 
\bibitem{Motome2010} Y. Motome, K. Nakamikawa, Y. Yamaji, and M. Udagawa: Phys. Rev. Lett. {\bf 105} (2010) 036403. 
\bibitem{Shiba1990} H. Shiba and P. Fazekas: Prog. Theor. Phys. Supplement {\bf 101} (1990) 403.
\bibitem{Sorella1998} S. Sorella: Phys. Rev. Lett. {\bf 80} (1998) 4558.
\bibitem{Tahara2008} D. Tahara and M. Imada: J. Phys. Soc. Jpn. {\bf 77} (2008) 114701, and references therein.
\bibitem{note} The 24-site cluster has a slightly different shape from that used in Ref.~7. 
\bibitem{note2} The two results will coincide as $N \to \infty$ 
when the system realizes a three-sublattice order with zero momentum.
\bibitem{Moessner2001} R. Moessner and S. L. Sondhi: Phys. Rev. B {\bf 63} (2001) 224401. 
\end{thebibliography}
\end{document}